\documentclass[aps,pre,superscriptaddress,twocolumn,longbibliography]{revtex4}
\usepackage{graphicx,epstopdf,url}
\usepackage[colorlinks=true,urlcolor=blue,citecolor=blue,linkcolor=blue,urlcolor=blue]{hyperref}
\usepackage[active]{srcltx}
\usepackage{multirow}
\begin{document}
\title{
Stabilization of  molecular hydrogen-bonded chains by carbon nanotubes
}

\author{Alexander V. Savin}

\affiliation{Semenov Institute of Chemical Physics, Russian Academy of Sciences,
Moscow 119991, Russia}
\affiliation{
Plekhanov Russian University of Economics, Moscow 117997, Russia
}
\email[]{asavin@chph.ras.ru}

\author{Yuri S. Kivshar}
\affiliation{Nonlinear Physics Center, Research School of Physics, Australian National University, Canberra ACT 2601, Australia}
\email[]{yuri.kivshar@anu.edu.au}

\begin{abstract}
We study numerically nonlinear dynamics of  several types of molecular systems composed of hydrogen-bonded chains placed inside carbon nanotubes with open edges. We demonstrate that carbon nanotubes provide a stabilization mechanism for quasi-one-dimensional molecular chains  via the formation of their secondary structures. In particular, a polypeptide chain (Gly)$_N$ placed inside a carbon nanotube can form of a stable helical chain ($3_{10}$, $\alpha$, $\pi$ and $\beta$-helix) with parallel chains of hydrogen-bonded peptide groups. A chain of hydrogen fluoride molecules  can form hydrogen-bonded zigzag chain. We reveal that in such geometries the hydrogen-bonded chains may remain stable even at $T=500$~K.  Thus, our results suggest that the use of carbon nanotubes with encapsulated hydrogen fluoride molecules may support high proton conductivity operating at high temperatures.
\end{abstract}

\maketitle

\maketitle

\section{Introduction}
\label{Introduction}

One of the research activities of David Campbell has been devoted to the study of nonlinear waves and solitons in discrete systems~\cite{david0,david00}, including the pioneering results on nonlinear dynamics of polyacetylene $[(CH)_x]$ ~\cite{david1}. Many of the discrete nonlinear models are inspired by the physics of molecular systems. Realistic molecular systems have a complex multi-component structure, consisting of nonlinearly interacting subsystems (electronic, excitonic, phononic, ionic, proton). In such systems, stable propagation of nonlinear waves (such as polarons, solitons, kinks, etc.) can be observed when nonlinear effects compensate for linear dispersion. As a special example, highly efficient energy and charge transfer takes place in molecular chains with hydrogen bonds. 

Molecular systems with hydrogen bonds are characterised by a relatively low stabilisation energy (0.12 - 0.40~eV per hydrogen bond), which ensures their high dynamic mobility.  Such structures include $\alpha$-helices and $\beta$-sheets of proteins~ \cite{Pauling51,Pauling51a}, DNA double helices, organic crystals of acetamides (e.g, acetanilide (CH$_3$CONHC$_6$H$_5$)$_x$~\cite{Careri84,Eilbeck84}), crystals of aramides (Kevlar crystals~\cite{Chowdhury18}), proton transport channels in cell membranes~ \cite{Nagle78, Nagle83,Kreuer96,Nagamani11}, crystals of alcohols, hydrogen halides~ \cite{Jansen87,Sprinborg88}, and some other systems.

However, many of the molecular systems studied theoretically are restricted by the assumption of low dimensionality, so they become unstable being place into three-dimensional space. Also, molecular complexes with hydrogen-bonded chains become unstable at high temperatures (at $T>100^\circ$C all hydrogen bonds are destroyed).
In this paper, we study numerically the dynamics of quasi one-dimensional molecular systems with hydrogen-bonded chains that can be stabilized being placed inside open carbon nanotubes. In particular, we demonstrate that a polypeptide chain (Gly)$_N$ may take the form of a stable helical chain ($3_{10}$, $\alpha$, $\pi$ and $\beta$-helix) with parallel chains of hydrogen-bonded peptide groups. A chain of hydrogen fluoride molecules  placed inside a nanotube of small radius can form hydrogen-bonded zigzag chain $\cdots$FH$\cdots$FH$\cdots$FH supporting proton transport. In such molecular systems, the hydrogen-bonded chains retain their structure and the interaction with the nanotube lead to their additional stabilization.  For some molecular structures the hydrogen-bonded chains will remain stable even at $T=500$~K. 

\section{Nonlinear dynamics of molecular hydrogen-bonded chains}

The presence of hydrogen-bonded chains of hydroxyl groups in the molecular system 
\begin{equation}
{\cdots}{\rm O{-}H}{\cdots}{\rm O{-}H}{\cdots}{\rm O{-}H}{\cdots}{\rm O{-}H}{\cdots}{\rm O{-}H}
\label{f2}
\end{equation}
ensures its high proton conductivity in the direction of these chains \cite{Zundel00}.
Proton transport across cell membranes takes place through protein proton channels and passes through hydrogen bond chains (\ref{f2}) formed by amino acid residues containing the OH hydroxyl group (serine, threonine, tyrosine) \cite{Nagle78}. 
In the bacteriorhodopsin molecule, the hydrogen bond chain is formed by tyrosine residues contained in the seven transmembrane $\alpha$-helical segments of the molecule \cite{Merz81}. 
Hydrogen-bonding chains (\ref{f2}) act as proton "wire"\ and provide an efficient pathway for rapid proton transfer \cite{Fillaux02}.

\begin{figure*}[tb]
\begin{center}
\includegraphics[angle=0, width=0.8\linewidth]{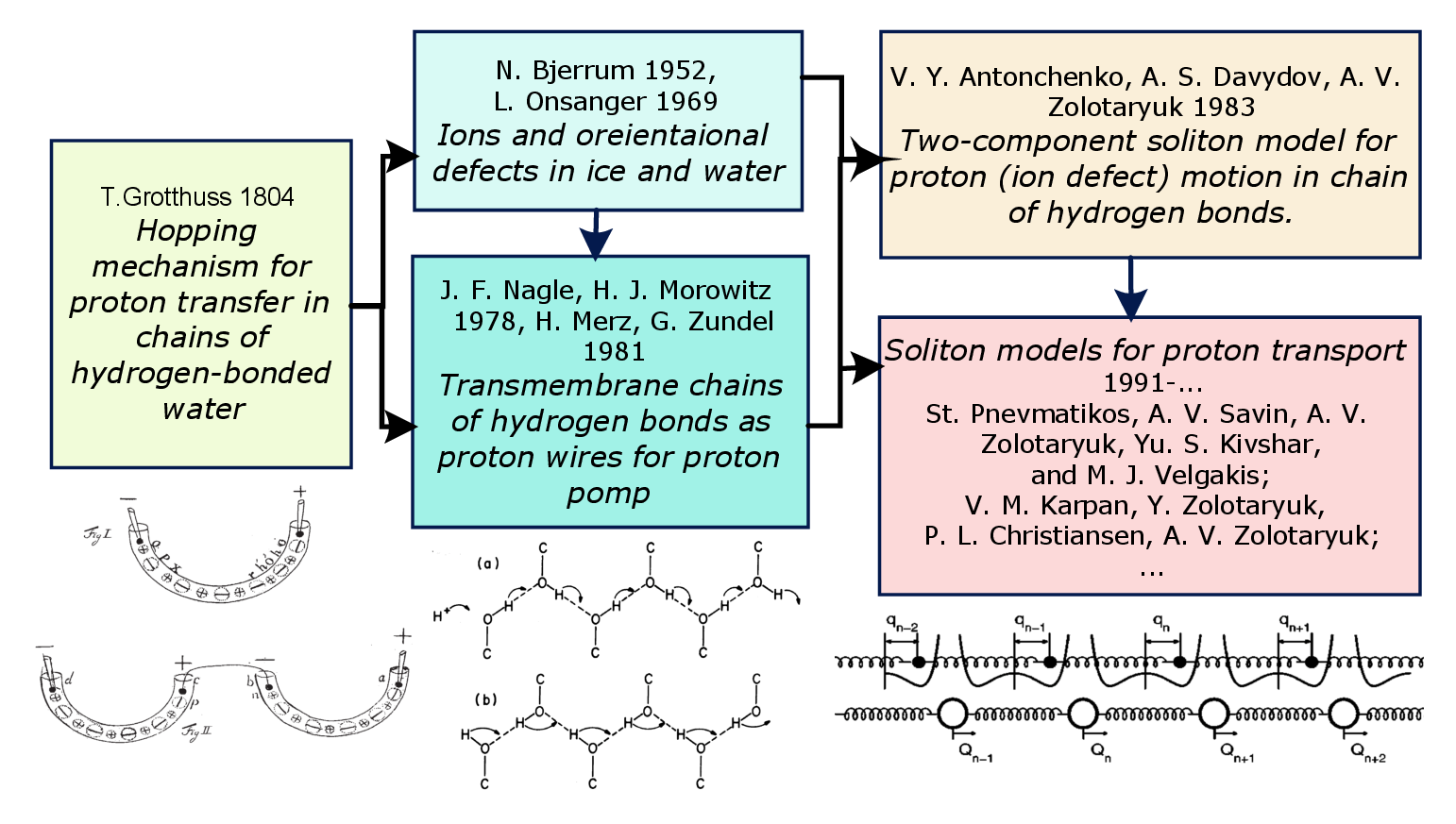}
\end{center}
\caption{\label{PrtnSoliton}\protect
Milestones of the nonlinear theory of hydrogen-bonded molecular chains and soliton-based proton transport.
}
\end{figure*}

The idea of proton transport through the hydrogen bond chains of water molecules was proposed by Theodore von Grotthuss in 1804 \cite{Marx06}. 
According to modern ideas, proton transport in water and ice takes place along the hydrogen bonding lattice and is divided into two phases: the passage of the ionic defect (H$^{+}$) (proton transport between water molecules involving the exchange of a covalent bond between O and H with a hydrogen bond) and the passage of the orientational defect (Bjerrum defect), which returns the hydrogen bonding chain to its initial state \cite{Bjerrum52}. 
Proton transport along the zigzag chain of hydrogen bonds occurs in the form of migration of two types of defects: ionic and orientational \cite{Nagle83,Onsager69}.
Defect migration can be described as the motion of topological solitons 
\cite{Antonchenko83,Pnevmatikos91,Karpan04,Pang11,Ndjike21}.
The creation and development of the theory of proton soliton transport in the literature is shown schematically in Fig. \ref{PrtnSoliton}.

Based on the helical structure of protein \cite{Pauling51,Pauling51a} and the theory of wave excitations (excitons) of molecular crystals \cite{Davydov51,Moffitt56,Davydov68,Davydov69} A.S. Davydov and N.I. Kislukha in 1973 \cite{Davydov73} (see also \cite{Davydov76,Davydov79,Davydov81}), proposed the theory of soliton energy transport along the alpha-helix of protein molecules.
The alpha-helix of a protein is stabilised by three parallel chains of hydrogen-bonded HNCO peptide groups (PGs): 
\begin{equation}
{\rm H{-}N{-}N{-}C{=}O}{\cdots}{\rm H{-}N{-}N{-}C{=}O}{\cdots}{\rm H{-}N{-}N{-}C{=}O}{\cdots},
\label{f1}
\end{equation}
where the lines represent valence bonds and the dots represent hydrogen bonds.
According to Davydov's theory, these chains can transport energy released during the hydrolysis of the
adenosine triphosphate (ATP) molecule in the form of self-localised states of vibrations of amide-I double valence bonds C=O, which are part of PG. 
The self-localisation of the amid-I vibration is caused by its non-linear interaction with longitudinal deformations of the hydrogen bond chains (\ref{f1}).
Using this model, it was shown that in $\alpha$-helical protein molecules nonlinear collective self-localised excitations can propagate without energy loss and shape change, which were later called {\it Davydov solitons}~\cite{Hyman81,Scott82,Lomdahl85,Scott85,Scott92}.


In the first papers by Davydov et al. the continuum approximation was used and the $\alpha$-helix was treated as a continuous medium. 
The discrete model was used to numerically model the dynamics of Davydov solitons in the work of Scott {\it et al.}~\cite{Hyman81,Scott82}. 
These works have led to an increased interest in the Davydov model
\cite{Lomdahl85a,Kuprievich85,Brizhik88,Savin93,Brizhik93,Zolotaryuk95,Brizhik04}.  The Davydov model is still actively used today, see Refs.~\cite{Georgiev20,Georgiev22,Cruzeiro16,Cruzeiro20,Cruzeiro22}.

The presence of hydrogen bonding chains (\ref{f1}) also allows for the possibility of external electron transfer in $\alpha$-helical proteins in the form of an electrosoliton (a bound state of a localised electron with the region of helix deformation resulting from the interaction of the electron with the PG chains) \cite{Davydov79a,Davydov84,Davydov91}.

Practically all molecular complexes with hydrogen bonding chains are no longer stable at high temperatures (at $T>100^\circ$C all hydrogen bonding chains are destroyed).
This complicates the use of such molecular systems in nanotechnology.
In this work it will be shown that the use of complexes of such molecular systems with carbon nanotubes
can significantly increase the stability of the hydrogen bond chains.
For this purpose, quasi one-dimensional molecular systems ($\alpha$-helix of protein, zigzag chain of hydrogen fluoride molecules) will be placed in carbon nanotubes of a certain radius.
In this case, the hydrogen bonding chains will retain their structure and the interaction with the nanotube 
will lead to their additional stabilisation, which will allow them to be used for energy and charge transfer over a wider temperature range.
In such composite systems the hydrogen bonding chains will remain stable even at $T=500$~K.
\begin{figure*}[tb]
\begin{center}
\includegraphics[angle=0, width=0.8\linewidth]{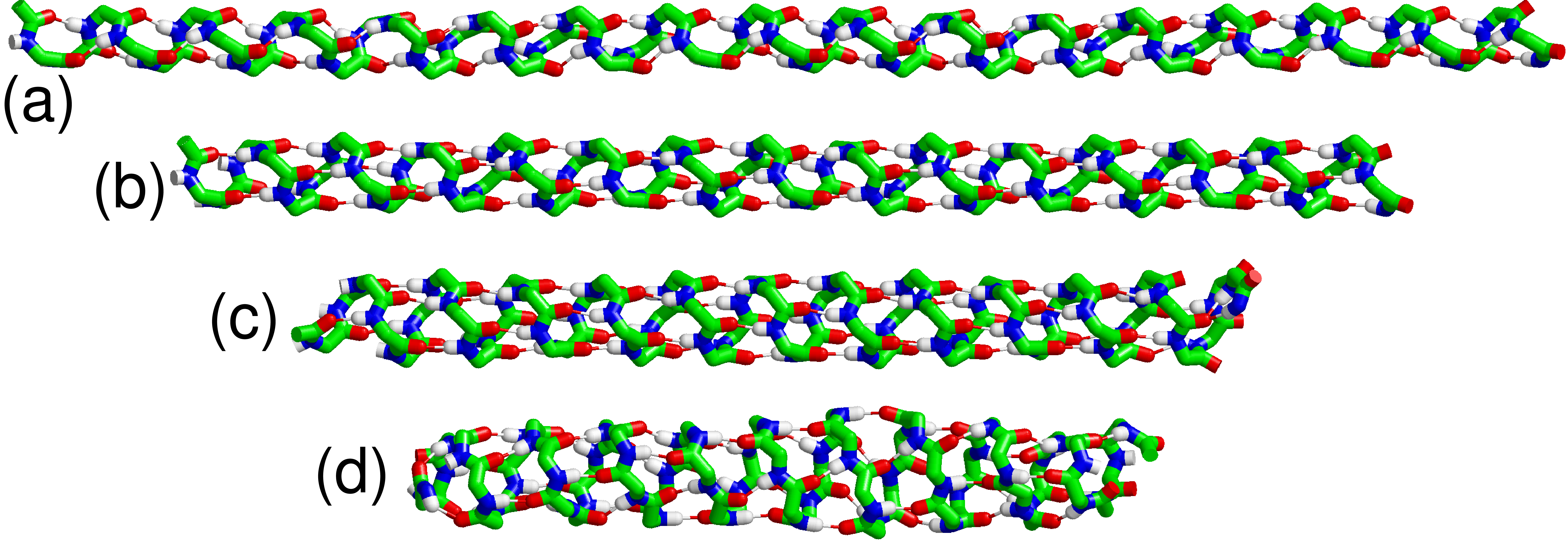}
\end{center}
\caption{\label{fig01}\protect
Helical structures of the peptide chain (Gly)$_{52}$: 
(a) $3_{10}$-helix (stationary structure energy $E=16.26$~eV, 
hydrogen bond energy $E_{\rm hb}=0. 175$~eV, length $L=9. 66$~nm); 
(b) $\alpha$ helix ($E=12.00$, $E_{\rm hb}=0.258$, $L=7.64$); 
(c) $\pi$ helix ($E=11.32$, $E_{\rm hb}=0.268$, $L=6.01$); 
(d) $\beta$ helix ($E=10.03$, $E_{\rm hb}=0.268$, $L=4.67$).
Carbon atoms are shown in green, nitrogen atoms in blue, oxygen atoms in red and hydrogen atoms in white.
Hydrogen bonds are shown as thin lines. Hydrogen atoms of $\alpha$-carbon atoms are not shown.
}
\end{figure*}

\section{Helices of proteins}

To analyse helical structures, consider a polypeptide chain (Gly)$_N$ consisting of $N$ peptide groups.
The chemical formula of such a chain is  
\begin{equation}
{\rm C}_\alpha{\rm H}_3{\rm-HNCO{-(}C}_\alpha{\rm H}_2{\rm{-HNCO-)}}_{N-2}{\rm C}_\alpha{\rm H}_2{\rm-H}_2{\rm NCO},
\label{f3}
\end{equation}
where two hydrogen atoms are attached to each $\alpha$-carbon atom (in order not to leave free valences, three hydrogen atoms are attached to the first $\alpha$-carbon atom and one hydrogen atom is attached to the nitrogen atom of the last peptide group).

When modelling a polyglycine chain (\ref{f3}), it is convenient to consider $\alpha$-carbon atoms with attached hydrogen atoms as one united atom $C_\alpha$. 
In this case each chain link will consist of 5 atoms C$_\alpha$, C, O, N, H with coordinates 
$\{ {\bf u}_{n,i}\}_{i=1}^5$, where $n$ is the link number of the polypeptide chain 
and $i$ is the atom number.
The atomic masses are $M_{n,1}=14m_p$, $M_{n,2}=12m_p$, $M_{n,3}=16m_p$, $M_{n,4}=14m_p$, $M_{n,5}=m_p$ (for end links $M_{1,1}=15m_p$, $M_{N,4}=15m_p$), where $m_p=1. 6603\times 10^{-27}$kg -- the mass of a proton.
 To describe the interatomic interaction, we will use the AMBER general force field for organic molecules (version 2.1, April 2016) \cite{Amber}.

The Hamiltonian of a polypeptide chain has the form 
\begin{equation}
H=\sum_{n=1}^N\sum_{i=1}^5\left[\frac12(M_{n,i}\dot{\bf u}_{n,i},\dot{\bf u}_{n,i}) +P(\{ {\bf u}_{n,i}\}_{n=1,i=1}^{N,~5})\right],
\label{f4}
\end{equation} 
where the first sum defines the kinetic energy 
(the three-dimensional vector ${\bf u}_{n,i}=(x_{n,i},y_{n,i},z_{n,i})$ defines the coordinates of the $i$-th atom of the $n$-th chain link), and the last sum -- the potential energy of the chain found using the AMBER force field.

To find the stationary state of the chain, we need to numerically solve the minimum potential energy problem
\begin{equation}
P\rightarrow \min: \{ {\bf u}_{n,i}\}_{n=1,i=1}^{N,~5}.
\label{f5}
\end{equation}
The numerical solution of the problem (\ref{f5}) showed that the polypeptide chain (\ref{f3}) has four stationary states as a helix chain: $3_{10}$-helix, $\alpha$-helix, $\pi$-helix and $\beta$-helix -- see Fig.~\ref{fig01}. 
Let us assume that the peptide groups $n$ and $k$ ($n<k$) form a hydrogen bond if the non-valent interaction energy of the corresponding chain links is
$$
E_{n,k}=U(\{{\bf u}_{n,i}\}_{i=1}^5,\{ {\bf u}_{k,j}\}_{j=1}^5)<-0.16~{\rm eV}.
$$ 
Each helix is characterised by the steady state energy $E$, the number of hydrogen bond chains $N_{\rm sp}$, the number of hydrogen bonds $N_{\rm hb}$, the average bond energy 
$$ 
E_{\rm hb}=-\sum_{l=1}^{N_{\rm hb}}E_{n_l,k_l}/N_{\rm hb}
$$
and the length of the helix (the distance between the end alpha carbon atoms 
$L=|{\bf u}_{N,1}-{\bf u}_{1,1}|$).
\begin{table}[tb]
\caption{
Energy $E$, number of hydrogen bond chains $N_{\rm sp}$, number of hydrogen bonds $N_{\rm hb}$, average hydrogen bond energy $E_{\rm hb}$ and chain end distance $L$ for different stationary states (conformations) of the polypeptide chain (Gly)$_{73}$.
\label{tb1}
}
\begin{center}
\begin{tabular}{c|ccccc}
~~conformation~~&~$E$~(eV)~&~$N_{\rm sp}$~&~$N_{\rm hb}$~&~$E_{\rm hb}$~(eV)~&~$L$~(nm)~\\
 \hline\hline
 $3_{10}$-helix  &  22.56   &  2            &  71          &  0.1704       &  13.59 \\
$\alpha$-helix &  16.22   &  3            &  70          &  0.2583       &  10.79 \\
 $\pi$-helix   &  14.93   &  4            &  69          &  0.2695       &   8.46 \\
 $\beta$-helix &  14.00   &  5            &  70          &  0.2560       &   5.94 \\
 globule       &  13.33   &  -            &  60          &  0.2341       &   0.57 \\  
\hline \hline
\end{tabular}
\end{center}
\end{table}

The chain of $N=73$ peptide groups was used to analyse the helical structures.
The values of $E$, $N_{\rm sp}$, $N_{\rm hb}$, $E_{\rm hb}$, and $L$ for different stationary states (conformations) of the polypeptide chain (Cly)$_{73}$ are given in Table \ref{tb1}.
As can be seen from the table, the most energetically unfavourable structure is the $3_{10}$ helix.
Such a helix is highly stretched, it has two chains of weakened hydrogen bonds -- see Fig.~\ref{fig01}.
Further more favourable in energy are the $\alpha$, $\pi$, $\beta$ helix and globule structures.

Let us consider the dynamics of helical structures. 
To do this, we place the polypeptide chain in a Langevin thermostat of temperature $T$ and simulate the dynamics of the system numerically during the time $t=100$~ns.
To do this, we numerically integrate the system of Langevin equations
\begin{eqnarray}
\label{f6}
M_{n,i}\ddot{\bf u}_{n,i}=-\frac{\partial H}{\partial {\bf u}_{n,i}}-\Gamma M_{n,i}\dot{\bf u}_{n,i}-\Xi_{n,i},\\
~n=1,...,N,~i=1,...,5, \nonumber
\end{eqnarray}
where $\Gamma=1/t_r$ is the friction coefficient, $\Xi_{n,i}=\{\xi_{n,i,k}\}_{k=1}^3$ is 3-dimensional
 vector of normally distributed random Langevin forces with the following correlations:
 $$
 \langle \xi_{n_1,i,k}(t_1)\xi_{n_2,j,l}(t_2)\rangle
 =2M_{n_1,i}k_BT\Gamma\delta_{n_1,n_2}\delta_{ij}\delta_{kl}\delta(t_1-t_2).
 $$
Here $M_{n,i}$ is mass of $i$-th atom of $n$-th segment of polypeptide chain,
$k_B$ is Boltzmann constant, $T$ is temperature of the Langevin thermostat, numbers 
$n_1,n_2=1,...,N$, $i,j=1,...,5$, $k,l=1,2,3$. The parameter $t_r=10$~ps characterizes
the intensity of energy exchange between molecular system and the thermostat.

The equations of motion Eq. (\ref{f6}) are solved numerically using the velocity Verlet method \cite{Swope82}.
A time step of 1 fs is used in the simulations, since further reduction of the time step has
no appreciable effect on the results.

Let us numerically integrate the system of equations of motion (\ref{f6}) with the initial condition corresponding to the stationary state of the polypeptide chain.
After the dynamics of the molecular system reaches the steady state, we will find
the time averages of the system energy $\bar{E}(T)$, the number
of hydrogen bonds $\bar{N}_{\rm hb}(T)$ and chain length $\bar{L}(T)$.

\begin{figure}[tb]
\begin{center}
\includegraphics[angle=0, width=1.0\linewidth]{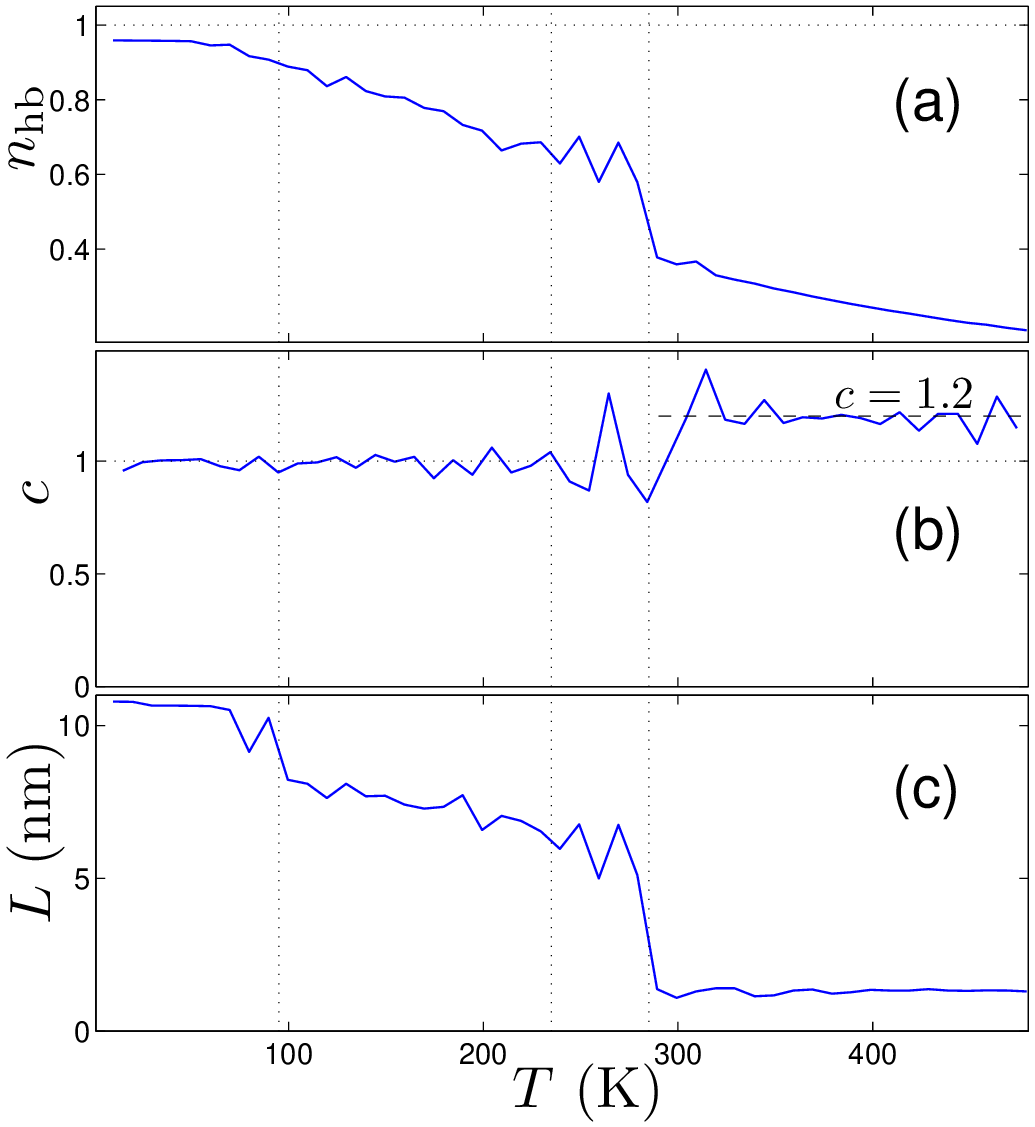}
\end{center}
\caption{\label{fig02}\protect
Dependence of (a) the normalised number of hydrogen bonds $n_{\rm hb}$, 
(b) the dimensionless heat capacity $c$ and (c) the chain end spacing $\bar{L}$ 
for a polypeptide chain (Gly)$_{73}$ originally shaped as a $\alpha$-helix.
The vertical lines indicate the temperature values of 95, 235 and 285K.
}
\end{figure}

The state of the system can be conveniently characterized by its dimensionless heat capacity
\begin{equation}
c=\frac{1}{15Nk_B}\frac{d\bar{E}(T)}{dT},
\label{f7}
\end{equation}
the normalized number of hydrogen bonds $n_{\rm hb}=\bar{N}_{\rm hb}(T)/N$ and chain length $\bar{L}(T)$.
The dependence of these quantities on temperature is shown in Fig. \ref{fig02}.

The $3_{10}$-helix is the most unstable. 
At temperature $T=10$~K it changes to $\pi$-helix.
The alpha helix of the chain remains stable at $T<T_1=95$~K, 
at $T_1<T<T_2=235$~K the $\alpha$-helix changes into $\pi$-helix, 
and at $T_2<T<T_3=285$~K into $\beta$-helix. 
At $T\ge T_3$, the chain transforms into a globule shape -- see Fig.~\ref{fig02}.
The $\pi$-helix remains stable at $T<T_2$, at $T=T_2$ it changes into a $\beta$-helix.
The beta helix is stable at $T<T_3$, at $T>T_3$ the chain adopts the globule form.
The transitions between the forms of the helix occur without changing the heat capacity 
of the molecular system, i.e. these are second order phase transitions.
The increase in heat capacity occurs only when the chain transitions to the globular form.
The temperature $T=T_3$ can be considered as the melting point of the $\beta$-helix.
The transition of a helix into a globule is first order phase transition,
here the dimensionless heat capacity increases from $c=1$ to $c=1.2$.

Thus, in an isolated polyglycine peptide chain, regular chains of hydrogen bonds (\ref{f1}) can only exist at low temperatures $T<285$~K. 

\section{Conformations of a peptide chain inside a carbon nanotube}

Polypeptide molecules tend to self-insertion into carbon nanotubes with open ends. 
Inside the nanotube, polypeptide atoms are able to fully realise non-valent interactions with its surface \cite{Liu05,Xiu13,Zhang14}.
This is particularly significant for hydrophobic polypeptides such as (Gly)$_N$, (Ala)$_N$, (Val)$_N$, (Ile)$_N$.
Let us determine what conformations the polypeptide chain (Gly)$_N$ inside a single-walled carbon nanotube with chirality index $(n,m)$ (CNT$_{(n,m)}$) can take.

To describe dynamics of CNT we present the nanotube Hamiltonian in the form
\begin{equation}
H_{\rm c}=\sum_{n=1}^{N_{\rm c}}\left[\frac12M_{n,0}(\dot{\bf v}_n,\dot{\bf v}_n)+Q_n\right],
\label{f8}
\end{equation}
where $N_{\rm c}$ is the number of atoms in the CNT, $M_{n,0}$ is the mass of $n$-th carbon atom,
${\bf v}_n=\{v_{n,i}(t)\}_{i=1}^3$ is the radius vector of $n$-th atom at the time $t$.
The term $Q_n$ describes the interaction energy of the atom with index $n$ with neighboring atoms.

To describe the carbon-carbon valence interactions, let us use a standard set molecular dynamics 
potentials \cite{Savin10,Savin17}. We consider a hydrogen-terminated nanotube with open ends, where
edge atoms correspond to the molecular group CH. We will consider such a group as a single effective
particle at the location of the carbon atom. Therefore, in our model of open CNT we take the mass of
atoms inside the nanotube as $M_{n,0}=12m_p$, and for the edge atoms as $M_{n,0}=13m_p$.

The valence bond between two neighboring carbon atoms $n$ and $k$ can be described by the Morse potential
\begin{equation}
U_1({\bf v}_n,{\bf v}_k)=\epsilon_1\{\exp[-\alpha(r_{nk}-r_1)]-1\}^2,
\label{f9}
\end{equation}
where $r_{nk}=|{\bf v}_n-{\bf v}_k|$, $\epsilon_1=4.9632$~eV is the valence bond energy, 
and $r_1=1.418$~\AA~ is the equilibrium valence bond length.

The valence angle deformation energy between three adjacent carbon atoms $n$, $k$, and $l$ 
can be described by the potential
\begin{equation}
U_2({\bf v}_n,{\bf v}_k,{\bf v}_l)=\epsilon_2(\cos\varphi-\cos\varphi_0)^2,
\label{f10}
\end{equation}
where $\cos\varphi=({\bf v}_n-{\bf v}_k,{\bf v}_l-{\bf v}_k)/r_{nk}r_{kl}$, 
and $\varphi_0=2\pi/3$ is the equilibrium valent angle. Parameters $\alpha=1.7889$~\AA$^{-1}$ and
$\epsilon_2=1.3143$~eV can be found from the small-amplitude oscillations spectrum of 
the graphene sheet \cite{Savin08}. Valence bonds between four adjacent carbon atoms $n$, $m$, $k$, 
and $l$ constitute torsion angles, whose potential energy can be defined as
\begin{equation}
U_3(\phi)=\epsilon_3(1-\cos\phi), 
\label{f11}
\end{equation}
where $\phi$ is the corresponding torsion angle ($\phi=0$ is the
equilibrium value of the angle) and $\epsilon_3=0.499$~eV.
A detailed discussion about the choice of the interatomic potential parameters can be found in \cite{Savin10}.

We describe the interaction of polypeptide chain atoms with nanotube atoms using Lennard-Jones potentials with parameters from the AMBER force field.
\begin{table}[tb]
\caption{
Values of the nanotube radius $R$, the helix type of the polypeptide chain, the energy $E$ of the stationary state of the complex (Gly)$_{73}{\in}$CNT$_{(n,m)}$, the interaction energy $E_{\rm i}$ of the polypeptide chain with the nanotube, and the average hydrogen bonding energy $E_{\rm hb}$ for different nanotubes with chirality index $(n,m)$ (nanotube length $L_{\rm cnt}=16. 1$~nm).
\label{tb2}
}
\begin{center}
\begin{tabular}{cc|cccc}
 ~$(n,m)$~ & ~$R$~(\AA)~     &~~helix& ~~$E$~(eV)~~& ~~$-E_{\rm i}$~(eV)~~ &~~$E_{\rm hb}$~(eV)~~ \\
 \hline\hline
 (12,0)               & 4.74                  & $3_{10}$ & 221.53   & 39.16  & 0.2756 \\ \hline
 (7,7)                & 4.76                  & $3_{10}$ & 219.25   & 39.66  & 0.2744 \\ \hline
\multirow{2}{*}{(13,0)}&\multirow{2}{*}{5.13} & $3_{10}$ & 204.16   & 37.33  & 0.2005 \\
                       &                     & $\alpha$ & 203.88   & 33.39  & 0.2711 \\ \hline
  (8,8)                &    5.43             & $\alpha$ & 188.93   & 34.06  & 0.2618 \\ \hline 
  (14,0)               &    5.52             & $\alpha$ & 186.57   & 33.65  & 0.2584 \\ \hline 
\multirow{2}{*}{(15,0)}&\multirow{2}{*}{5.91}& $\alpha$ & 178.93   & 28.74  & 0.2314 \\ 
                       &                     & $\pi$    & 175.29   & 30.87  & 0.2705 \\ \hline
\multirow{3}{*}{(9,9)} &\multirow{3}{*}{6.11}& $\alpha$ & 175.68   & 25.53  & 0.2481 \\
                       &                     & $\pi$    & 170.79   & 29.12  & 0.2641 \\
                       &                     & $\beta$  & 171.71   & 30.33  & 0.2588 \\ \hline
\multirow{3}{*}{(16,0)}&\multirow{3}{*}{6.29}& $\alpha$ & 172.18   & 24.25  & 0.2558 \\
                       &                     & $\pi$    & 167.92   & 27.03  & 0.2614 \\
                       &                     & $\beta$  & 167.64   & 28.63  & 0.2532 \\ \hline
\multirow{3}{*}{(17,0)}&\multirow{3}{*}{6.68}& $\alpha$ & 164.70   & 21.74  & 0.2568 \\
                       &                     & $\pi$    & 161.99   & 22.92  & 0.2652 \\
                       &                     & $\beta$  & 157.05   & 27.68  & 0.2708 \\ \hline
\multirow{3}{*}{(10,10)}&\multirow{3}{*}{6.78}& $\alpha$& 162.77   & 21.06  & 0.2555 \\
                       &                     & $\pi$    & 160.16   & 22.15  & 0.2658 \\
                       &                     & $\beta$  & 155.55   & 26.46  & 0.2700 \\ \hline
\multirow{3}{*}{(18,0)}&\multirow{3}{*}{7.07}& $\alpha$ & 157.64   & 19.89  & 0.2587 \\
                       &                     & $\pi$    & 155.23   & 20.83  & 0.2659 \\
                       &                     & $\beta$  & 152.06   & 23.31  & 0.2653 \\ \hline
\multirow{3}{*}{(11,11)}&\multirow{3}{*}{7.46}& $\alpha$& 150.77   & 18.48  & 0.2570 \\
                       &                     & $\pi$    & 148.85   & 19.33  & 0.2660 \\
                       &                     & $\beta$  & 146.56   & 20.24  & 0.2636 \\
\hline \hline
\end{tabular}
\end{center}
\end{table}
\begin{figure*}[tb]
\begin{center}
\includegraphics[angle=0, width=0.9\linewidth]{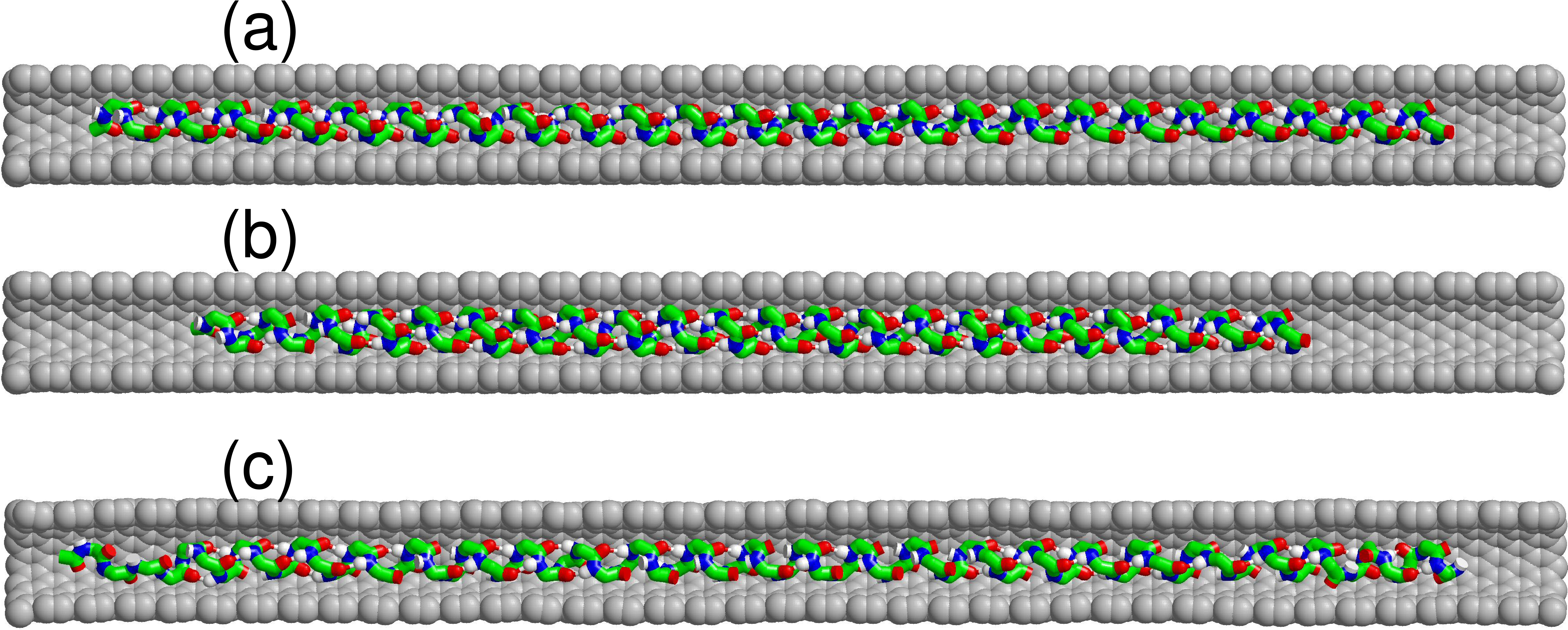}
\end{center}
\caption{\label{fig03}\protect
State of the polypeptide chain (Gly)$_{73}$ located in the nanotube (13,0) with open edges: 
(a) $3_{10}$-helix ($T=10$~K), (b) $\alpha$-helix at $T=10$~K, and (c) $T=300$K (the peptide chain is $3_{10}$-helix).
For ease of review, only one half of the nanotube atoms is shown (nanotube atoms are shown in grey).
The length of the nanotube is $L_{\rm cnt}=16.14$~nm, the number of carbon atoms is $N_{\rm c}=1976$.
}
\end{figure*}

The Hamiltonian of the molecular system (Gly)$_N{\in}$CNT will have the form 
\begin{eqnarray}
H=\sum_{n=1}^{N_{\rm c}}\frac12(M_{n,0}\dot{\bf v}_n,\dot{\bf v}_n)+
\sum_{n=1}^N\sum_{i=1}^5\frac12(M_{n,i}\dot{\bf u}_{n,i},\dot{\bf u}_{n,i})
\nonumber\\
+E(\{ {\bf v}_n\}_{n=1}^{N_{\rm c}},\{ {\bf u}_{n,i}\}_{n=1,i=1}^{N,~5}),~~~
\label{f12}
\end{eqnarray}
where the first summand defines the kinetic energy of the carbon nanotube 
($N_{\rm c}$ is the number of atoms of the nanotube), 
the second -- the kinetic energy of the polypeptide chain, 
the third -- the interaction energy of the atoms of the system.

To find the stationary state of the two-component system (Gly)$_N{\in}$CNT it is necessary to solve numerically the problem on the minimum of the interaction energy
\begin{equation}
E\rightarrow\min: \{ {\bf v}_n\}_{n=1}^{N_{\rm c}},\{ {\bf u}_{n,i}\}_{n=1,i=1}^{N,~5}.
\label{f13}
\end{equation}

Let us take the polypeptide chain (Gly)$_{73}$ and place it inside a nanotube of length $L_{\rm cnt}=16.1$~nm.
The numerical solution of the problem (\ref{f13}) shows that the polypeptide chain inside the nanotube takes the form of a helix, the type of which depends on the radius of the nanotube $R$ 
-- see Table. \ref{tb2}.
Thus at $R<5$~\AA~, the chain inside the nanotube can take only one stationary state -- helix 3$_{10}$.
Two stationary states of the chain helix $3_{10}$ and $\alpha$-helix are possible inside the nanotube (13,0) -- see Fig.~\ref{fig03} (a),(b). 
The $\alpha$-helix can exist in all nanotubes with radius $R>5.1$~\AA, $\pi$-helix -- in nanotubes with $R>5.9$~\AA, $\beta$-helix -- in nanotubes with $R>6.1$~\AA. 
Note that in nanotubes with $R<5$~\AA~, hydrogen bonds become strong for 3$_{10}$ helix,  the strongest interaction between the polypeptide chain and the nanotube occurs here (the smaller the radius of the nanotube, the greater the interaction energy).

Let us consider the dynamics of the (Gly)$_{73}{\in}$CNT complex. 
To do this, we place the nanotube in a Langevin thermostat of temperature $T$ (the thermalisation of the polypeptide chain inside the CNT will occur through its interaction with the nanotube) and numerically simulate the dynamics of the molecular complex during the time $t=20$~ns. To do this, we numerically integrate the system of Langevin equations
\begin{eqnarray}
M_{n,0}\ddot{\bf v}_n=-\frac{\partial H}{\partial {\bf v}_n}-\Gamma M_{n,0}\dot{\bf v}_n-\Theta_n,
\label{f14} \\
n=1,...,N_{\rm c} \nonumber\\ 
M_{n,i}\ddot{\bf u}_{n,i}=-\frac{\partial H}{\partial {\bf u}_{n,i}}, \label{f15}\\
n=1,...,N,~i=1,...,5, \nonumber
\end{eqnarray}
where $\Theta_n=\{\eta_{n,k}\}_{k=1}^3$ is 3-dimensional vector of normally distributed random
Langevin forces with the following correlations:
$$
\langle\eta_{n_1,k}(t_1)\eta_{n_2,l}(t_2)\rangle=2M_{n_1,0}k_BT\Gamma\delta_{n_1,n_2}\delta_{kl}\delta(t_1-t_2).
$$ 

We numerically integrate the system of equations of motion (\ref{f14}), (\ref{f15}) with the initial condition corresponding to the solution of the minimum problem (\ref{f13}).
After the dynamics of the molecular complex reaches the steady state, we will find the time averages
of the complex energy $\bar{E}(T)$, the number of hydrogen bonds $\bar{N}_{\rm hb}(T)$ and peptide
chain length $\bar{L}(T)$.

\begin{figure}[tb]
\begin{center}
\includegraphics[angle=0, width=1.0\linewidth]{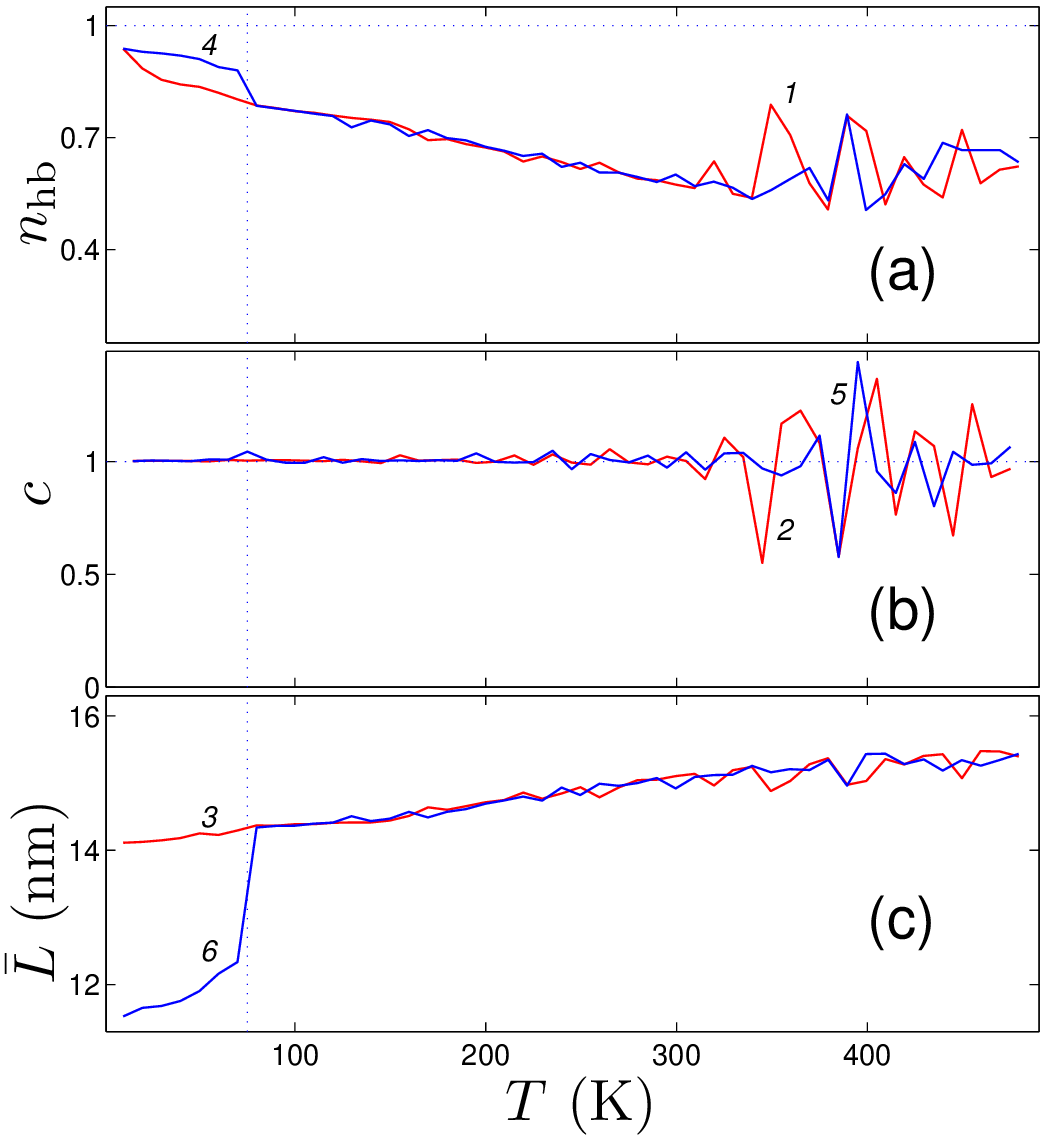}
\end{center}
\caption{\label{fig04}\protect
Dependence of (a) the normalised number of hydrogen bonds $n_{\rm hb}$, 
(b) the dimensionless heat capacity $c$ and 
(c) the distance between the ends of the peptide chain $\bar{L}$ 
for the molecular complex (Gly)$_{73}{\in}$CNT$_{(13,0)}$.
Curves 1, 2, 3 show dependencies for the $3_{10}$ helix chain, 
curves 4, 5, 6 -- for the $\alpha$ helix chain.
The length of the CNT is $L_{\rm cnt}=16.14$~nm, the number of atoms $N_{\rm c}=1976$, 
at $T=75$K there is a transition from $\alpha$-helix to $3_{10}$-helix.
}
\end{figure}

\begin{figure}[tb]
\begin{center}
\includegraphics[angle=0, width=1.0\linewidth]{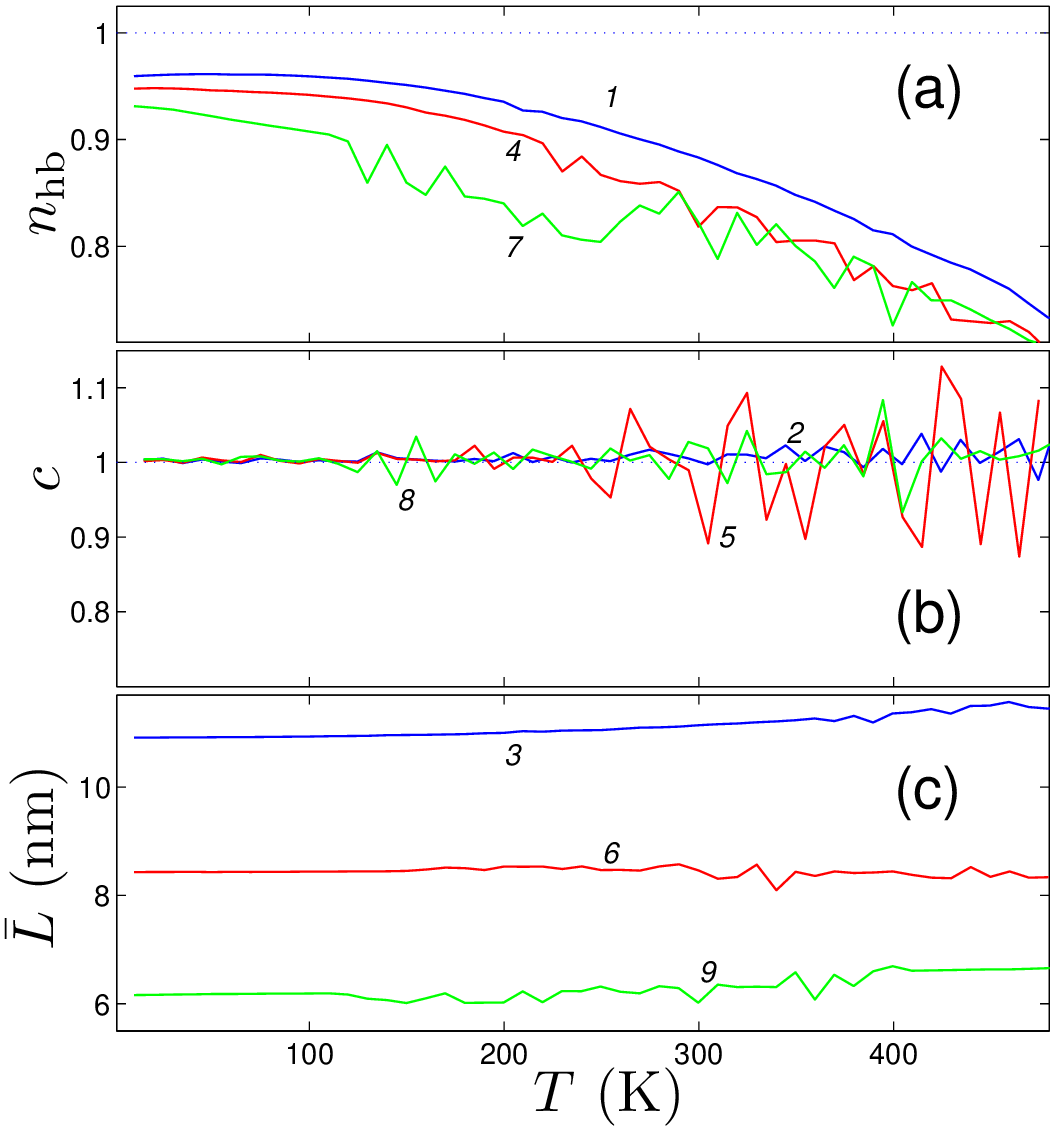}
\end{center}
\caption{\label{fig05}\protect
Dependence of (a) the normalized number of hydrogen bonds $n_{\rm hb}$,
(b) the dimensionless heat capacity $c$ and (c) 
the distance between the ends of the peptide chain $\bar{L}$
for the molecular complex (Gly)$_{73}{\in}$CNT for nanotube (8,8) (curves 1, 4, 7),
(9,9) (curves 2, 5, 8) and (10,10) (curves 3, 6, 9).
In nanotube (8,8), the peptide chain is $\alpha$-helix, 
in (9,9) is $\pi$-helix, in (10,10) is $\beta$-helix.
}
\end{figure}

The state of the molecular complex can be characterized by its dimensionless heat capacity
\begin{equation}
c=\frac{1}{3(N_{\rm c}+5N)k_B}\frac{d\bar{E}(T)}{dT},
\label{f16}
\end{equation}
the normalized number of hydrogen bonds $n_{\rm hb}=\bar{N}_{\rm hb}(T)/N$ and peptide chain length
$\bar{L}(T)$. The dependence of this quantities on temperature for different nanotubes is shown
in Fig.~\ref{fig04} and \ref{fig05}.  
\begin{figure}[tb]
\begin{center}
\includegraphics[angle=0, width=1.0\linewidth]{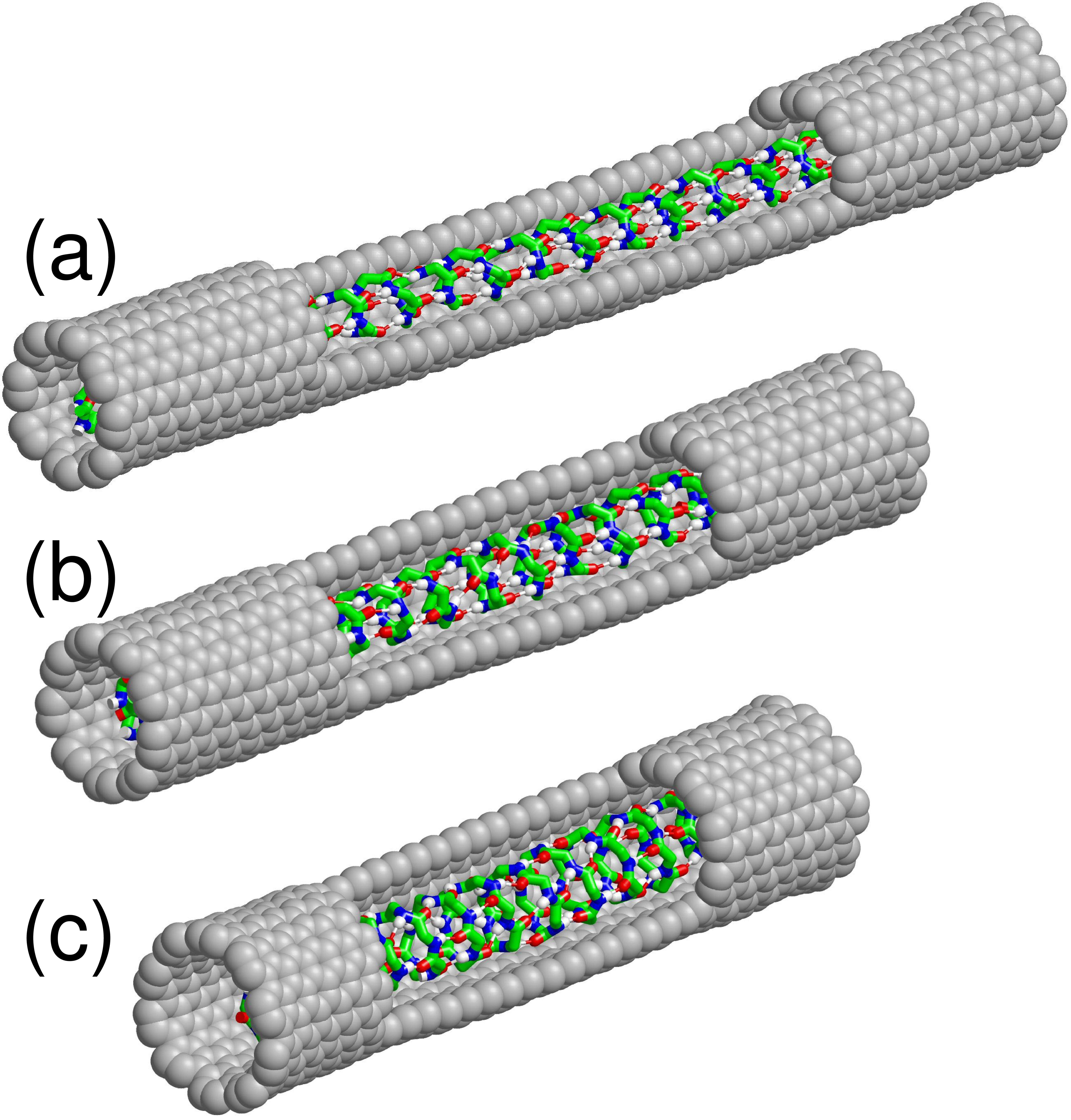}
\end{center}
\caption{\label{fig06}\protect
Configuration of peptide chain (Gly)$_{73}$ in (a) carbon nanotube (8,8) ($L_{\rm cnt}=11$~nm), 
(b) nanotube (9,9) ($L_{\rm cnt}=8.8$) and (c) nanotube (10,10) ($L_{\rm cnt}=7.3$).
The temperature is $T=300$K. In nanotube (8,8) the chain is in the form of a $\alpha$-helix, 
in nanotube (9,9) in the form of a $\pi$-helix, and in (10,10) in the form of a $\beta$-helix.
For convenience, some carbon atoms of the nanotubes are not shown.
}
\end{figure}

Numerical integration of the system of equations of motion (\ref{f14}), (\ref{f15}) showed that at $T<500$~K there is only one basic helix type for each nanotube. 
For CNT$_{(13,0)}$ the $3_{10}$ helix will be the basic helix. 
Here, at $T<75$~K, two forms of helix can exist: $3_{10}$ and $\alpha$ helix, and at $T>75$~K, only the $3_{10}$ helix can exist (thermal fluctuations lead to the transition of the $\alpha$ helix to the $3_{10}$ helix) -- see Fig.~\ref{fig03} and Fig.~\ref{fig04}. 
The $3_{10}$ helix exists at all temperatures $T<500$~K.
As the temperature increases, only partial breaking of the edge hydrogen bonds occurs, leading to a decrease in $n_{\rm hb}$ and an increase in the peptide chain length $\bar{L}$.
The dimensionless heat capacity of the (Gly)$_{73}{\in}$CNT$_{(13,0)}$ complex remains constant at $c=1$.

Note that for an isolated peptide chain, the $3_{10}$ helix form is the most unstable.
Even at $T=10$~K, thermal fluctuations lead to a rapid transition to the $\pi$-helix, and at $T=285$~K the chain melts (the chain changes from helical to molten globule form).

In nanotube (8,8) the ground state of the peptide chain will always be $\alpha$-helix, in CNT (9,9) -- $\pi$-helix and in CNT (10,10) -- $\beta$-helix.
The appearance of these helical states of the peptide chain in nanotubes with chirality indices (8,8), (9,9) and (10,10) is shown in Fig. \ref{fig06}. 
All these helixes are stable to thermal fluctuations at $T<500$~K. 
With increasing temperature there is only a small decrease in the number of hydrogen bonds -- see Fig.~\ref{fig05}~(a).
At the same time, the dimensionless heat capacity always remains equal to one and the length of the helix is almost unchanged -- see Fig.~\ref{fig05}~(b) and (c).

Thus, placing a polypeptide chain in a nanotube of small radius will stabilise its helical shape. 
For the simplest polypeptide chain (Gly)$_N$, the most stable to thermal fluctuations will be the $\alpha$-helix lying inside the nanotube (9,9).
Here the $\alpha$-helix is conserved at all temperatures $T<500$~K. 
Note that such stabilisation of the helical shape is impossible for an isolated polypeptide chain.
Note that the optimal nanotube radius for helix stabilisation depends on the size of the amino acid residues. 
For larger Ala, Val, Ile residues, the best stabilisation will occur in nanotubes of larger radius.

\section{Hydrogen-bonded chains of hydrogen fluoride molecules}

Experimental and theoretical studies of water-filled carbon nanotubes have shown that water molecules can enter the interior of the open nanotubes and form extended hydrogen bond chains 
\cite{Hummer01,Dellago03,Kofinger11,Mendonca19,Druchok23}.
Small-diameter nanotubes with a chain of water molecules inside can act as proton-conducting channels \cite{Chen13,Ma20}.

Carbon nanotubes are hydrophobic and the presence of water molecules inside them is not energetically favourable. 
Inside the nanotube, water molecules cannot organise a three-dimensional lattice of hydrogen bonds.
For example, inside a CNT$_{(6,6)}$ (diameter $D=0.80$~nm) only one chain of water molecules can be found, in which each molecule participates in the formation of two hydrogen bonds (in the bulk phase each molecule participates in the formation of four bonds) \cite{Hanasaki08}.

Hydrogen fluoride molecules FH can also form zigzag chains of hydrogen bonds (FH)$_\infty$: 
\begin{equation} 
{\cdots}{\rm F{-}H}{\cdots}{\rm F{-}H}{\cdots}{\rm F{-}H}{\cdots}{\rm F{-}H}{\cdots}{\rm F{-}H},
\label{f18}
\end{equation}
which are similar to zigzag chains of hydroxyl groups (\ref{f2}).
At temperature $T<189.6$~K, hydrogen fluoride has a crystal structure formed by parallel flat hydrogen bond chains \cite{Atoji54}.
Note that a separate zigzag chain of hydrogen bonds (\ref{f18}) becomes unstable in three-dimensional space. 
The chain breaks down into cyclic chains of small size, the most stable being chains of 6 molecules in the shape of a regular hexagon \cite{Orabi20}.
Hydrogen bond chains (\ref{f18}) can be stabilised by placing them inside the CNT, i.e. forming a (FH)$_\infty{\in}$CNT structure.

In each chain (\ref{f18}), one hydrogen fluoride molecule participates in the formation of two hydrogen bonds.
Analysis of the interaction of the F--H${\cdots}$F--H dimer with the CNT shows that it is energetically more favourable for the hydrogen fluoride dimer to be inside the nanotube than outside \cite{Roztoczynska16}.
Therefore, (FH)$_N$ chains can easily be placed inside open CNTs to form stable (FH)$_N{\in}$CNT structures ($N$ -- number of hydrogen fluoride molecules). 
Keeping hydrogen fluoride inside nanotubes may help to avoid problems associated with its high toxicity \cite{Gtari18}.
 
Let us simulate molecular complexes (FH)$_N{\in}$CNT.
We describe the valence bond deformation of the two-atom molecule FH by a harmonic potential 
\begin{equation}
V(\rho)=\frac12K(\rho-\rho_0)^2,
\label{f19}
\end{equation}
where $\rho$ and $\rho_0=0.929$~\AA~ are the current and equilibrium bond lengths, $K=444.3$~N/m is the bond stiffness.

The interaction of FH molecules is conveniently described by the potential 12-6-1 \cite{Cournoyer84,Orabi20}
\begin{equation}
U=\sum_{i=1}^3\sum_{j=1}^3\kappa q_iq_j/r_{ij}+4\epsilon[(\sigma/r)^{12}-(\sigma/r)^6],
\label{f20}
\end{equation}
where the first sum defines the Coulomb interaction between systems of point charges approximating the charge distribution in a pair of interacting FH molecules ($r_{ij}$ -- the distance between the charge $q_i$ of the first molecule and the charge $q_j$ of the second molecule), $r$ -- the distance between the centres of the fluorine atoms, the coefficient $\kappa=14. 400611$~eV\AA/$e^2$.
Two positive charges $q_2$, $q_3$ are on atoms F and H, the negative charge $q_1=-(q_2+q_3)$ is at a distance $r_1$ from atom F on the interval [FH]. 
If for the potential (\ref{f20}) does not fix the position of the two charges on the atoms of the molecule, the potential will be given by seven parameters: two charges, three distances $\{r_i\}_{i=1}^3$, specifying the position of the charges on the line connecting atoms F and H, and two parameters $\epsilon$ and $\sigma$ of the Lenard-Jones potential.
\begin{figure}[tb]
\begin{center}
\includegraphics[angle=0, width=1.0\linewidth]{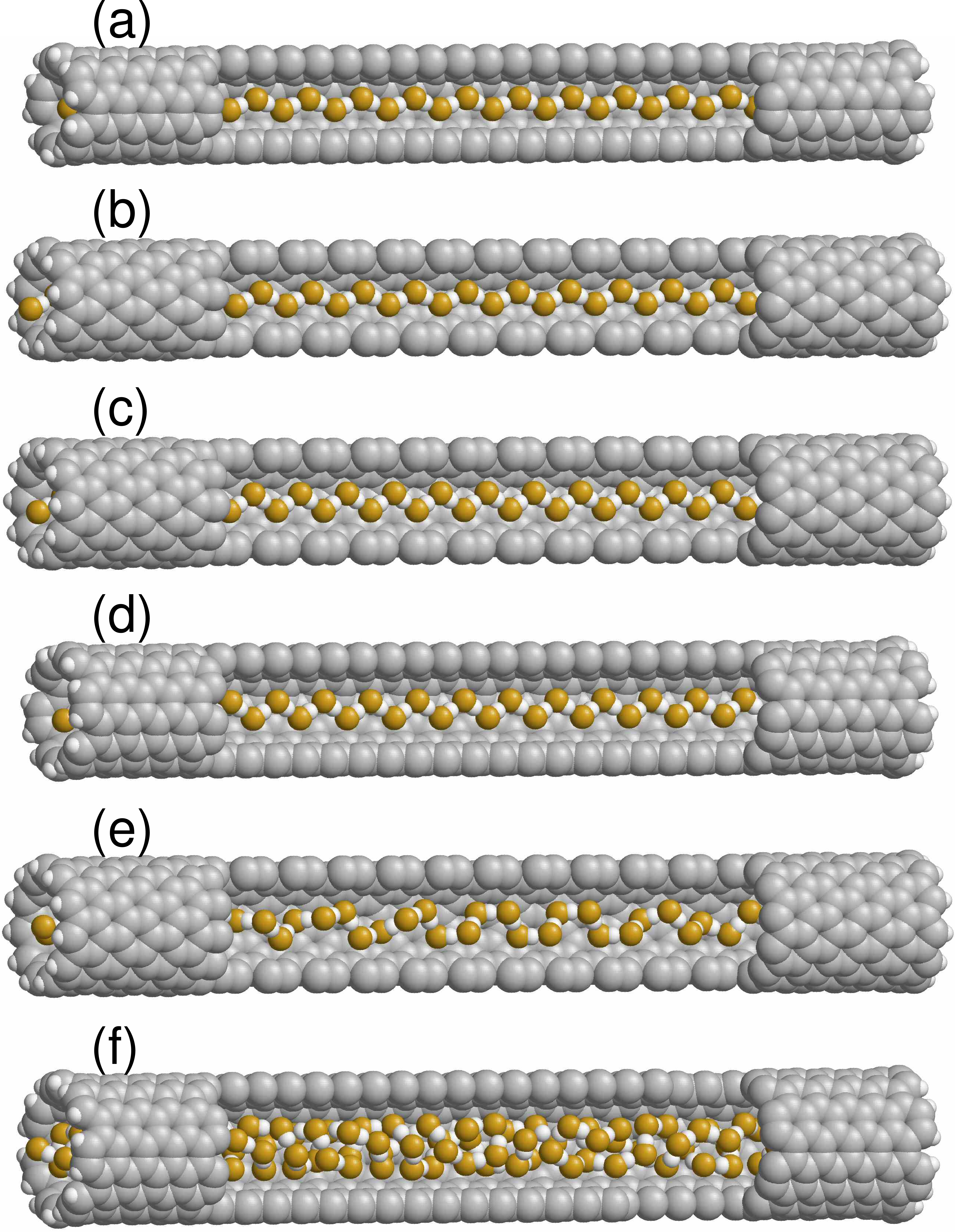}
\end{center}
\caption{\label{fig10}\protect
Stationary state of the hydrogen bond chain (FH)$_N$ inside an open single-walled CNT with chirality index $(n,m)$:
(a) (FH)$_{32}{\in}$CNT$_{(5,5)}$ (nanotube chemical formula C$_{600}$H$_{20}$);
(b) (FH)$_{35}{\in}$CNT$_{(9,0)}$ (C$_{648}$H$_{18}$);
(c) (FH)$_{38}{\in}$CNT$_{(10,0)}$ (C$_{720}$H$_{20}$);
(d) (FH)$_{36}{\in}$CNT$_{(6,6)}$ (C$_{720}$H$_{24}$);
(e) (FH)$_{44}{\in}$CNT$_{(11,0)}$ (C$_{792}$H$_{22}$);
(f) (FH)$_{106}{\in}$CNT$_{(7,7)}$ (C$_{840}$H$_{28}$).
For ease of review, some atoms of the nanotubes are not shown.
}
\end{figure}

The relative simplicity and a sufficiently large number of free parameters of the 12-6-1 potential make it very convenient for modelling the dynamics of multi-molecular complexes of two-atom FH molecules.
The FH molecule usually participates in the formation of two strong hydrogen bonds, polarising significantly in the process. 
We will assume that when the valence bond length $r=|{\rm FH}|$ changes, the charge arrangement distances from the centre of the fluorine atom also change proportionally. 
To model the  dynamics of the hydrogen bond chain (\ref{f18}) it is best to use the potential (\ref{f20}) with the parameters
\begin{eqnarray}
q_1=-0.6397e,~~q_2=0.6159e,~~q_3=0.0238e,
\nonumber\\
r_1=0.2577\rho/\rho_0,~~r_2=0.9356\rho/\rho_0,~~r_3=-1.6237\rho/\rho_0,
\nonumber\\
\epsilon=0.0079778~{\rm eV},~~\sigma=2.837109~{\rm \AA},
\nonumber
\end{eqnarray}
where $\rho$ and $\rho_0$ are the actual and equilibrium lengths of the F-H valence bond, $e$ is the electron charge (dimension $[r_i]=$\AA). 
With this set of parameters, the FH molecule will have the correct values for the dipole and quadrupole moments 
$$
\mu=\sum_{i=1}^3q_ir_i,~~Q= \sum_{i=1}^3 q_i(r_i-r_m)^2
$$ 
($r_m$ -- the distance from the centre of gravity of the molecule to the centre of the fluorine atom).
The energy and configuration of ground-state of the dimer (FH)$_2$ will best agree with the experimental data \cite{Dyke72} and the results of the quantum calculations \cite{Nemuchin92}.

Consider open armchair and zigzag nanotubes with hydrogen atoms attached to the edge carbon atoms -- see Fig.~\ref{fig10}. 
To find the ground state (FH)$_N{\in}$CNT it is necessary to solve numerically the problem on the minimum of potential energy of the system (\ref{f13}). 
By choosing the initial position of the molecules, we can obtain all possible stationary states of the two-component system.
At the position of the chain (FH)$_N$ along the outer surface of the nanotube we obtain a flat zigzag chain of hydrogen bonds with pitch $a=|{\rm FF}|=2.49$~\AA~ and zigzag angle $\alpha=\angle{\rm FFF}=117^\circ$.
Torsion angle of the chain (dihedral angle formed by four consecutive fluorine atoms) $\phi=\angle{\rm FFFF}=180^\circ$.

\begin{table}[tb]
\caption{
Values of nanotube diameter $D$, pitch $a$ and angle $\alpha$ of zigzag and angle of torsion $\phi$ of hydrogen bonding chain (FH)$_\infty$ inside CNT$_{(n,m)}$.
\label{tb3}
}
\begin{center}
\begin{tabular}{c|cccc}
~~$(n,m)$~~ &~$D$~(\AA)~&~$a$~(\AA)~&~$\alpha$~($^\circ$)~&~$\phi$~($^\circ$)\\
 \hline\hline
 (5,5)  &  6.8   &  2.48   &  138   & 180  \\
 (9,0)  &  7.1   &  2.48   &  132   & 180  \\
 (10,0) &  7.8   &  2.47   &  114   & 180  \\
 (6,6)  &  8.1   &  2.47   &  113   & 180  \\
 (11,0) &  8.6   &  2.47   &  108   & 120  \\
 (7,7)  &  9.5   &  2.60   &   88   & 93  \\  
\hline \hline
\end{tabular}
\end{center}
\end{table}

The numerical solution of the minimum energy problem (\ref{f13}) showed that the hydrogen bond chain (\ref{f18}) retains a flat zigzag shape when placed inside nanotubes of diameter $D<0.85$~nm -- see Fig.~\ref{fig10} and Tab.~\ref{tb3}. 
The chain pitch $a$ is weakly dependent on the nanotube diameter, the chain zigzag angle $\alpha$ decreases monotonically with increasing diameter, the chain torsion angle $\phi=180^\circ$ at $D<0.85$~nm.
In the nanotube (11,0) ($D=0.86$~nm) the chain takes the form of a three-dimensional helix (torsion angle $\phi<180^\circ$) running along the nanotube surface.

\begin{figure}[tb]
\begin{center}
\includegraphics[angle=0, width=1.0\linewidth]{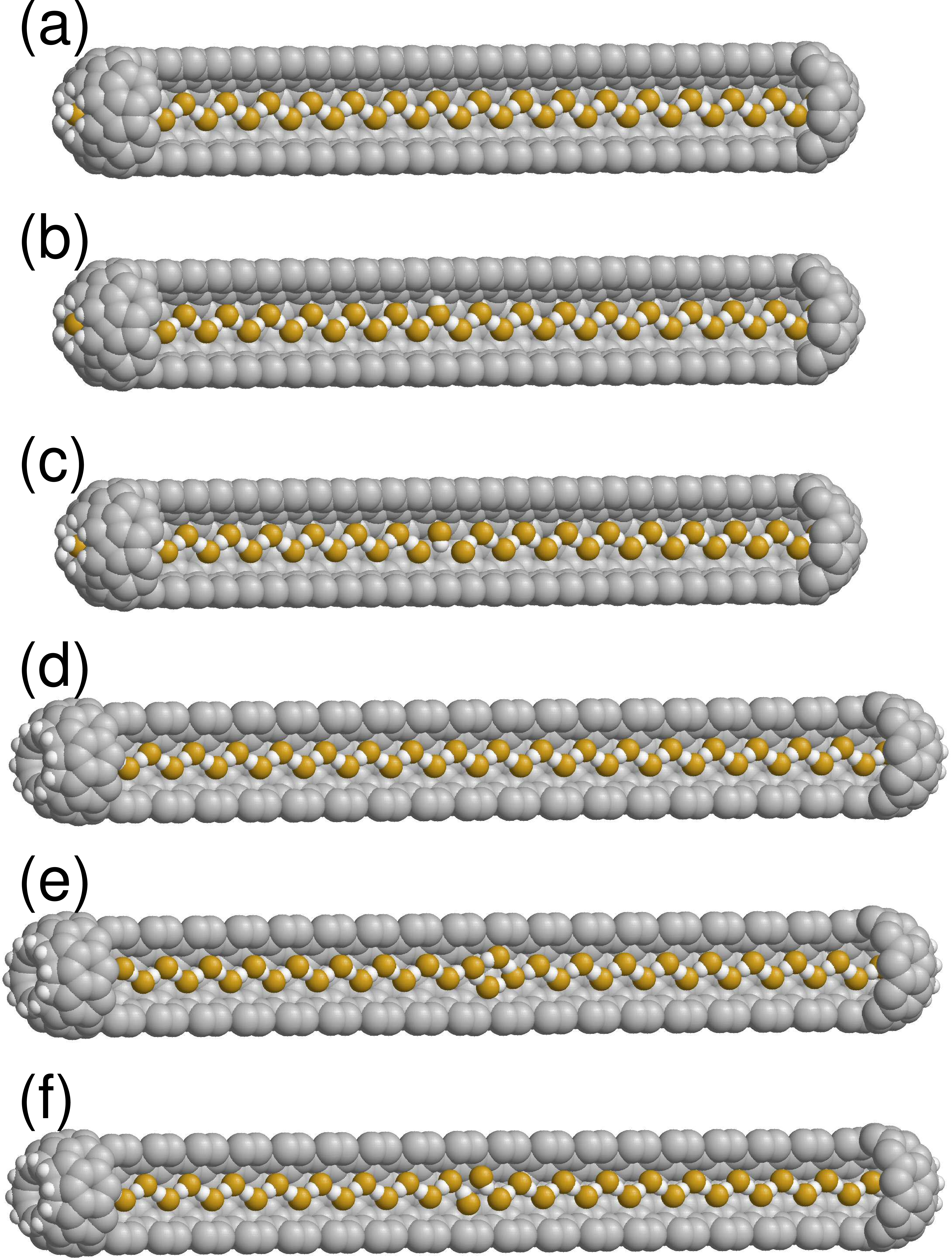}
\end{center}
\caption{\label{fig11}\protect
Stationary state of the hydrogen bond chain (FH)$_N$ inside CNT$_{(n,m)}$ with narrowed open edges:
(a) (FH)$_{36}$@CNT$_{(10,0)}$ (nanotube chemical formula C$_{430}$H$_{12}$), ground state;
(b) a chain with a positive orientational defect;
(c) a chain with a negative orientational defect;
(d) (FH)$_{38}$@CNT$_{(6,6)}$ (C$_{402}$H$_{20}$), ground state;
(e) a chain with a positive orientational defect;
(f) a chain with a negative orientational defect (defects are located in the centre of the chain).
For simplicity, some atoms of the nanotubes are not shown.
}
\end{figure}

The modelling shows that the hydrogen bond chain closest in structure to the hydrogen bond chain of the hydroxyl groups (\ref{f2}) is formed by hydrogen fluoride molecules inside nanotubes with chirality indices (6,6) and (10,0). We will therefore consider these complexes further.

To avoid the possibility of hydrogen fluoride molecules escaping from the nanotubes, it is better to narrow their open edges as shown in Fig. \ref{fig11}. 
The small diameter of the edge openings ($D=0.47$ and 0.42~nm) will prevent a large fluorine atom from leaving, but will preserve the possibility of protons leaving and entering.
So we get a chain of hydrogen bonds encapsulated in a carbon nanotube: (FH)$_N$@CNT.
Note that in 2016 endohedral fullerene HF@C60 was synthesised containing Buckminsterfullerene inside a molecule of hydrogen fluoride \cite{Krachmalnicoff16}.

Proton transport along the zigzag chain of hydrogen bonds proceeds by migration of two types of defects: ionic and orientational \cite{Nagle83,Onsager69}. 
In the first stage of transport, the proton moves as an ionic defect. 
There is a sequential movement of hydrogen atoms along the hydrogen bond lines from one fluorine atom to the neighbouring one, transferring the chain from the $\cdots$FH$\cdots$FH$\cdots$FH$\cdots$FH$\cdots$ to the $\cdots$HF$\cdots$HF$\cdots$HF$\cdots$HF$\cdots$. 
In the second stage, an orientation defect (sequential rotation of the FH chain molecules by the zigzag angle $\alpha$) must pass through the chain to bring the chain back to its original orientation.

Note that the used force field (\ref{f19}), (\ref{f20}) allows to model the formation and motion of orientational defects.

The chain of hydrogen bonds inside the nanotube is bistable, all molecules of the chain can be orientated either from left to right or from right to left. An orientation defect occurs when the molecules of the chain are directed in different directions in the first and second parts of the chain. 
When the molecules face each other, an excess density of protons appears in the defect localisation region, so such a defect is called a positive defect -- see Fig.~\ref{fig11} (b) and (e). 
When the molecules are directed away from each other in the defect localisation region, the proton density decreases, so such a defect is called negative -- see Fig.~\ref{fig11} (c) and (f).

The stationary orientational defects of the hydrogen bonding chain (FH)$_N$ inside CNT (6,6) and (10,0) are shown in Fig. \ref{fig11} (b, c, e, f). 
As can be seen from the figure, the orientational defects are localised at 3-4 chain links. 
In nanotube (6,6), the energy of positive defect (the difference of the energy of the chain with defect from the energy of the ground state of the chain) is $\Delta E_+=0.56$~eV, the energy of negative defect is $\Delta E_-=0.51$~eV. 
In the (10,0) nanotube, the orientational defect energies of the chain $\Delta E_+=0.06$~eV, $\Delta E_-=0.91$ eV.

To test the stability of a chain of hydrogen bonds (FH)$_N$ placed inside a CNT, a molecular dynamics simulation was carried out on a chain of $N=177$ FH molecules placed inside a CNT$_{(6,6)}$ with narrowed open edges.
With a length of $L=38.7$~nm, such a nanotube is presented by the formula C$_{3620}$H$_{20}$. 
A zigzag chain of hydrogen bonds (FH)$_{177}$ completely fills the internal cavity of the nanotube.
Such a nanotube of length $L=8.7$~nm is shown in Fig. \ref{fig11} (d).

Numerical modeling of the system of equations of motion (\ref{f14}), (\ref{f15}) showed that the chain of hydrogen bonds inside the nanotube, complex (FH)$_{177}$@CNT$_{(6,6)} $, at all considered temperatures $T<500$~K remains stable -- it always remains inside and maintains a zigzag shape.
At $T<400$~K, the chain of hydrogen bonds always remains in the initial ground state, and at $T>400$~K a positive orientational defect \cite{Savin20} can already appear at the right end of the chain and then move along it.

Simulation of the dynamics of a complex with an orientational defect in the internal chain of hydrogen bonds showed that the movement of the defects is thermally activated (at low temperatures the defects remain pinned) and occurs in the form of Brownian motion of the particle along the chain.
The higher the temperature, the higher the mobility of defects.
A positive orientation defect has higher mobility than a negative defect. 
A defect disappears when it reaches the end of the chain and the chain goes into a homogeneous ground state.

Thus, thermal vibrations do not lead to breaks in the chain of hydrogen bonds (FH)$_\infty$ inside the carbon nanotube, so the chain can always be used for proton transport.
When a proton is transferred along a chain of hydrogen bonds, the longest time is required for its reorientation after passing through the ionic defect.
This reorientation, the transition of the chain from the HF$\cdots$HF$\cdots$HF$\cdots$ state to the FH$\cdots$FH$\cdots$FH$\cdots$ state, occurs due to thermally activated passage along the chain of positive orientation defect.
Therefore, thermal vibrations not only do not interfere with the transport of protons along the chain of hydrogen bonds inside the nanotube, but they are its necessary condition.

\section{Conclusion}

Being inspired by the pioneering studies of David Campbell on nonlinear dynamics in molecular structures, here we have studied numerically the structure and dynamics of several types of hydrogen-bonded molecular chains placed inside caped or open carbon nanotubes. We have demonstrated that a polypeptide chain is stabilized by a carbon nanotube, and it may take the form of a helical structure with parallel chains of hydrogen-bonded peptide groups. A chain of hydrogen fluoride molecules  placed inside a nanotube of small radius can form hydrogen-bonded zigzag chain being highly suitable for the proton transport. In such molecular systems, the hydrogen-bonded chains retain their structure, being stabilized by the interaction with the nanotube.  For some molecular structures the hydrogen-bonded chains will remain stable even at $T=500$~K. Thus, the results of our numerical modeling suggest that the use of carbon nanotubes with encapsulated hydrogen fluoride molecules would allow to construct anhydrous molecular systems supporting high proton conductivity being capable of operating at high temperatures.

\begin{acknowledgments}
This work was supported by the
Australian Research Council (Grants Nos. DP200101168 and
DP210101292).  Computational facilities were provided by the Interdepartmental Supercomputer Center of the Russian Academy of Sciences.
\end{acknowledgments}


\newpage 

\begin{references}

\bibitem{david0}
D.K. Campbell, S. Flach, and Y.S. Kivshar,
Localizing energy through nonlinearity and discreteness,
Physics Today {\bf 57}, 43 (2004).
%
\bibitem{david00}
Y.S. Kivshar and D.K. Campbell,
Peierls-Nabarro potential barrier for highly localized nonlinear modes,
Physical Review E {\bf 48}, 3077 (1993).
\bibitem{david1}
K. Fesser, A.R. Bishop, and D.K. Campbell, 
Optical absorption from polarons in a model of polyacetylene
Physical Review B {\bf 27}, 4804 (1983). 
\bibitem{Pauling51}
L. Pauling, R. B. Corey, H. R. Branson. 
The structure of proteins; two hydrogen-bonded helical configurations of the polypeptide chain. 
Proc. Natl. Acad. Sci. USA {\bf 37}(4), 205-211 (1951).
\bibitem{Pauling51a}
L. Pauling, R. B. Corey. 
Atomic coordinates and structure factors for two helical configurations of polypeptide chains. 
Proc. Natl. Acad. Sci. USA  {\bf 37}(5), 235-240 (1951).
\bibitem{Careri84}
G. Careri, U. Buontempo, F. Galluzzi, A. C. Scott, E. Gratton, E. Shyamsunder. 
Spectroscopic evidence for Davydov-like solitons in acetanilide. 
Phys. Rev. B {\bf 30}(8), 4689-4702 (1984).
\bibitem{Eilbeck84}
J. C.Eilbeck, P. S. Lomdahl, A. C. Scott. 
Soliton structure in crystalline acetanilide. 
Phys. Rev. B. {\bf 30}(8), 4703-4712 (1984).
\bibitem{Chowdhury18}
S. C. Chowdhury, J. W. Gillespie Jr.
A molecular dynamics study of the effects of hydrogen bonds on mechanical
properties of Kevlar crystal.
Computational Materials Science {\bf 148} 286-300 (2018).
\bibitem{Nagle78}
J. F. Nagle, H. J. Morowitz. 
Molecular mechanisms for proton transport in membranes. 
Proc. Natl. Acad. Sci. USA {\bf 75}(1), 298-302 (1978).
\bibitem{Nagle83}
J. F. Nagle, S. Tristram-Nagle. 
Hydrogen bonded chain mechanisms for proton conduction and proton pumping.
J. Membrane Biol. {\bf 74}, 1-14 (1983).
\bibitem{Kreuer96}
K.-D. Kreuer. 
Proton Conductivity: Materials and Applications.
Chem. Mater. {\bf 8}(3), 610-641 (1996).
\bibitem{Nagamani11}
C. Nagamani, U. Viswanathan, C. Versek, M. T. Tuominen, S. M. Auerbach, S. Thayumanavan. 
Importance of dynamic hydrogen bonds and reorientation barriers in proton transport.
Chem. Commun. {\bf 47}, 6638-6640 (2011).
\bibitem{Jansen87}
 R. W. Jansen, R. Bertoncini, D. A. Pinnick, A. I. Katz, R. C. Hanson, O. F. Sonkey, M. O'Keeffe. 
 Theoretical aspects of solid hydrogen halides under pressure. 
 Phys. Rev. B {\bf 35}(18), 9830-9840 (1987).
\bibitem{Sprinborg88}
M. Sprinborg. 
Energy surfaces and electronic properties of hydrogen fluoride. 
Phys Rev. B {\bf 38}(2), 1483-1503 (1988).
\bibitem{Zundel00}
G. Zundel. 
Hydrogen Bonds with Large Proton Polarizability and Proton Transfer Processes 
in Electrochemistry and Biology.
Adv. Chem. Phys. {\bf 111}, 1 (2000)
\bibitem{Merz81}
H. Merz, G. Zundel. 
Proton conduction in bacteriorhodopsin VIA a hydrogen-bonded chain with large proton polarizability.
Biochem. Biophys. Res. Commun. {\bf 101}(2), 540-546 (1981).
\bibitem{Fillaux02}
F. Fillaux. 
The impact of vibrational spectroscopy with neutrons on our view of quantum dynamics in hydrogen
bonds and proton transfer.
J. Mol. Struct. {\bf 615}, 45-59 (2002).
\bibitem{Marx06}
D. Marx.
Proton Transfer 200 Years after von Grotthuss: 
Insights from Ab Initio Simulations.
Chem. Phys. Chem. {\bf 7}(9), 1848-1870 (2006).
\bibitem{Bjerrum52}
N. Bjerrum. 
Structure and Properties of Ice.
Science. {\bf 115}(2989), 385-390 (1952).
\bibitem{Onsager69}
L. Onsager. 
The Motion of Ions: Principles and Concepts.
Science {\bf 166}(3911), 1359 (1969).
\bibitem{Antonchenko83}
V. Y. Antonchenko, A. S. Davydov, and A. V. Zolotaryuk,
Solitons and proton motion in ice-like structures.
Phys. Status Solidi B {\bf 115}, 631 (1983).
\bibitem{Pnevmatikos91}
St. Pnevmatikos, A. V. Savin, A. V. Zolotaryuk, Yu. S. Kivshar, and M. J. Velgakis. 
Nonlinear transport in hydrogen-bonded chains: Free solitonic excitations. 
Phys. Rev. A. {\bf 43}(10), 5518-5536 (1991).
\bibitem{Karpan04}
V. M. Karpan, Y. Zolotaryuk, P. L. Christiansen, A. V. Zolotaryuk. 
Discrete kink dynamics in hydrogen-bonded chains: The two-component model.
Phys. Rev. E {\bf 70}, 056602 (2004).
\bibitem{Pang11}
X.-F. Pang, J.-F. Yu, and H.-J. Zeng.
The properties of proton conductivity along the hydrogen-bonded molecular systems 
with damping under influences of thermal perturbation and structure nonuniformity.
International Journal of Modern Physics B, {\bf 25}(01), 55-71 (2011).
\bibitem{Ndjike21}
M. B. T. Ndjike, A. S. T. Nguetcho, J. Li, J. M. Bilbault.
Interplay role between dipole interactions and hydrogen bonding on proton transfer dynamics.
Nonlinear Dyn {\bf 105}, 2619-2643 (2021).
\bibitem{Davydov51}
A. S. Davydov. Theory of the Absorption of Light in Molecular Crystals. Kiev:
Ukrainian Acad. Scien. (1951).
\bibitem{Moffitt56}
W. Moffitt. 
Optical Rotatory Dispersion of Helical Polymers. 
J. Chem. Phys. {\bf 25}(6), 467-478 (1956).
\bibitem{Davydov68}
A. S. Davydov, Theory of Molecular Excitons, Izd. Nauka, Moscow 1968;
A. S. Davydov. Theory of Moleculur Excitons. New York: Plenum Press. (1971).
\bibitem{Davydov69}
A. S. Davydov. Deformation of Molecular Crystals at Electronic Excitation. 
Phys. Stat. Sol. {\bf 36}, 211-219 (1969).
\bibitem{Davydov73}
A. S. Davydov, N. I. Kislukha. 
Solitary excitations in one-dimensional molecular chains. 
Phys. Stat. Sol. B {\bf 59}, 465-470 (1973).
\bibitem{Davydov76}
A. S. Davydov, N. I. Kislukha. 
Solitons in one-dimensional molecular chains. 
Phys. Stat. Sol. B {\bf 75}, 735-742 (1976). 
\bibitem{Davydov79}
A. S. Davydov. 
Solitons in molecular systems. 
Phys. Scr. {\bf 20}  387-394 (1979). 
\bibitem{Davydov81} 
A. S. Davydov. 
The role of solitons in the energy and electron transfer in one-dimensional molecular systems. 
Physica D {\bf 3}, 1-22 (1981).
\bibitem{Hyman81}
 J. M. Hyman, D. W. Mc Laughlin, A. C. Scott. 
On Davydov's alphahelix solitons.
Physica D {\bf 3}, 23-45 (1981).
\bibitem{Scott82}
A. C. Scott. 
Dynamics of Davydov solitons.
Phys. Rev. A {\bf 26}, 578-595 (1982).
\bibitem{Lomdahl85}
P. S. Lomdahl, S. P. Layne, I. J. Bigio. 
Solitons in biology.
Los Alamos Sci. {\bf 10}, 2-22 (1984).
\bibitem{Scott85}
A. C.  Scott. 
Davydov solitons in polypeptides. 
Phill. Trans. Roy. Soc. London A {\bf 315}, 423-436 (1985).
\bibitem{Scott92}
A. C. Scott. 
Davydov's soliton. 
Phys Rep {\bf 217}, 1-67 (1992).
\bibitem{Lomdahl85a}
P. S. Lomdahl, W. C. Kerr. 
Do Davydov solitons exist at 300K? 
Phys. Rev. Lett. {\bf 55}(11), 1235-8 (1985).
\bibitem{Kuprievich85}
V. A. Kuprievich
On autolocalization of the stationary states in a finite molecular chain.
Physica D {\bf 14}(3), 395-402 (1985).
\bibitem{Brizhik88}
L. S. Brizhik, Y. B. Gaididei, A. A. Vakhnenko, V. A. Vakhnenko. 
Soliton generation in semi-infinite molecular chains. 
Phys. Status Solidi B {\bf 146}(2), 605-12 (1988).
\bibitem{Savin93}
A. V. Savin, A. V. Zolotaryuk. 
Dynamics of the amide-I excitation in a molecular chain with thermalized acoustic and optical modes.
Physica D {\bf 68}(1), 59-64 (1993).
\bibitem{Brizhik93}
L. S. Brizhik. 
Soliton generation in molecular chains. 
Phys. Rev. B {\bf 48}(5),3142-4 (1993).
\bibitem{Zolotaryuk95}
A. V. Zolotaryuk, K. H. Spatschek, A. V. Savin. 
Bifurcation scenario of the Davydov-Scott self-trapping mode. 
Europhys. Lett. {\bf 31}(9),531-6 (1995).
\bibitem{Brizhik04}
L. Brizhik, A. Eremko, B. Piette, and W. Zakrzewski.
Solitons in $\alpha$-helical proteins.
Phys. Rev. E {\bf 70}, 031914 (2004).
\bibitem{Georgiev20}
D. D. Georgiev, J. F. Glazebrook.
Launching of Davydov solitons in protein $\alpha$-helix spines.
Physica E {\bf 124}, 114332 (2020).
\bibitem{Georgiev22}
D. D. Georgiev, J. F. Glazebrook.
Thermal stability of solitons in protein $\alpha$-helices.
Chaos, Solitons and Fractals {\bf 155}, 111644  (2022).
\bibitem{Cruzeiro16}
L. Cruzeiro.
The VES Hypothesis and Protein Conformational Changes.
Z. Phys. Chem. {\bf 230}(5-7), 743-776 (2016).
\bibitem{Cruzeiro20}
L. Cruzeiro.
The VES KM: a pathway for protein folding in vivo.
Pure Appl. Chem. {\bf 92}(1), 179-191 (2020).
\bibitem{Cruzeiro22}
L. Cruzeiro.
Knowns and unknowns in the Davydov model for energy transfer in proteins.
Fiz. Nizk. Temp. {\bf 48}, 1105-1126 (December 2022).
\bibitem{Davydov79a}
A. S. Davydov.
nfluence of electron-phonon interaction on the motion of an electron in a one-dimensional molecular system.
Theoretical and Mathematical Physics, 1979, Volume 40, Issue 3, Pages 831-840
DOI: https://doi.org/10.1007/BF01032070
\bibitem{Davydov84}
A. S. Davydov, A. V. Zolotaryuk. 
Subsonic and supersonic solitons in nonlinear molecular chains. 
Phys. Scripta. {\bf 30}, 426-430 (1984).
\bibitem{Davydov91}
A. S. Davydov, 
Solitons in Molecular Systems, (Reidel, Dord-recht, 1991).
\bibitem{Amber}
W. D. Cornell, W. P. Cieplak, C. I. Bayly, I. R. Gould, K. M. Merz, D. M.
Ferguson, D. C. Spellmeyer, T. Fox, J.W. Caldwell, and P. A. Kollman. 
A second generation force field for the simulation of proteins, nucleic acids,
and organic molecules. J. Am. Chem. Soc. {\bf 117}, 5179-5197 (1995).
\bibitem{Swope82}
W. C. Swope, H. C. Andersen, P. H. Berens, and K. R. Wilson. 
A computer simulation method for the calculation of equilibrium constants for
the formation of physical clusters of molecules: Application to small water clusters.
J. Chem. Phys. {\bf 76}, 637 (1982).
\bibitem{Liu05}
G. R. Liu, Y. Cheng, Dong Mi, and Z. R. Li.
A study on self-insertion of peptides into single-walled carbon nanotubes based 
on molecular dynamics simulation.
International Journal of Modern Physics {\bf 16}(08), 1239-1250 (2005).
\bibitem{Xiu13}
P. Xiu, Z. Xia and R. Zhou.
Small Molecules and Peptides Inside Carbon Nanotubes: Impact of Nanoscale Confinement.
DOI: 10.5772/51453
From the edited volume "Physical and Chemical Properties of Carbon Nanotubes",
Edited by Satoru Suzuki, 2013. 
\bibitem{Zhang14}
Z.-S. Zhang, Y. Kang, L.-J. Liang, Y.-C. Liu, T. Wu, Q. Wang.
Peptide encapsulation regulated by the geometry of carbon nanotubes.
Biomaterials {\bf 35}(5), 1771-1778 (2014).
\bibitem{Savin10}
A. V. Savin, Y. S. Kivshar, and B. Hu. 
Suppression of thermal conductivity in graphene nanoribbons with rough edges. 
Phys. Rev. B {\bf 82}, 195422 (2010).
\bibitem{Savin17}
A. V. Savin and Y. S. Kivshar. 
Phononic Fano resonances in graphene nanoribbons with local defects. 
Sci. Rep. {\bf 7}, 4668 (2017).
\bibitem{Savin08}
A. V. Savin and Y. S. Kivshar. 
Discrete breathers in carbon nanotubes. 
Europhys. Lett. {\bf 82}, 66002 (2008).
\bibitem{Hummer01}
G. Hummer, J. Rasaiah, J. Noworyta.
Water conduction through the hydrophobic channel of a carbon nanotube.
Nature {\bf 414}, 188 (2001).
\bibitem{Dellago03}
C. Dellago, M. M. Naor, G. Hummer. 
Proton Transport through Water-Filled Carbon Nanotubes.
Phys. Rev. Lett. {\bf 90}(10), 105902 (2003).
\bibitem{Kofinger11}
J. Kofinger, G. Hummer and C. Dellago. 
Single-File Water in Nanopores. 
Phys. Chem. Chem. Phys. {\bf 13}(34), 15403-15417 (2011). 
\bibitem{Mendonca19}
B. H. S. Mendonca, D. N. de Freitas, M. H. Kohler, R. J. C. Batista,
M. C. Barbosa, A. B. de Oliveira. 
Diffusion Behaviour of Water Confined in Deformed Carbon Nanotubes.
Physica A {\bf 517}, 491-498 (2019).
\bibitem{Druchok23}
M. Druchok, V. Krasnov, T. Krokhmalskii, T. C. E Bufalo, S. M. de Souza, O. Rojas, O. Derzhko. 
Toward a quasiphase transition in the single-file chain of water molecules: Simple lattice model.
J. Chem. Phys. {\bf 158}(10), 104304 (2023).
\bibitem{Chen13}
J. Chen,  X.-Z. Li,  Q. Zhang,  A. Michaelides  and  E. Wang.
Nature of proton transport in a water-filled carbon nanotube and in liquid water.
Phys. Chem. Chem. Phys. {\bf 15}, 6344-6349 (2013).
\bibitem{Ma20}
X. Ma, C. Li, A. B. F. Martinson, and G. A. Voth.
Water-Assisted Proton Transport in Confined Nanochannels.
Phys. Chem. C {\bf 124}(29), 16186-16201 (2020).
\bibitem{Hanasaki08}
I. Hanasaki, A. Nakamura, T. Yonebayashi, S. Kawano.
Structure and stability of water chain in a carbon nanotube.
J. Phys.: Condens. Matter {\bf 20}, 015213 (2008).
\bibitem{Atoji54}
M. Atoji, W. N. Lipscomb.
The crystal structure of hydrogen fluoride.
Acta Crystallographica {\bf 7}, 173 (1954).
\bibitem{Orabi20}
E. A. Orabi, J. D. Faraldo-Gomez.
A new molecular-mechanics model for simulations
of hydrogen fluoride in chemistry and biology.
J. Chem. Theory Comput. {\bf 16}(8), 5105-5126 (2020).
\bibitem{Roztoczynska16}
A. Roztoczynska, J. Koztowska, P. Lipkowski, W. Bartkowiak.
Hydrogen bonding inside and outside carbon nanotubes: HF dimer as a case study.
Phys. Chem. Chem. Phys. {\bf 18}, 2417 (2016).
\bibitem{Gtari18}
W. F. Gtari, B. Tangour. 
Interaction of HF, HBr, HCl and HI Molecules with Carbon Nanotubes.
Acta Chim. Slov. {\bf 65}, 289 (2018).
\bibitem{Cournoyer84}
M. E. Cournoyer, W. L. Jorgensen. 
An improved intermolecular potential function for simulations of liquid hydrogen fluoride.
Mol. Phys. {\bf 51}, 119 (1984).
\bibitem{Dyke72}
Radiofrequency and Microwave Spectrum of the Hydrogen Fluoride Dimer; a Nonrigid Molecule.
T. R. Dyke, B. J. Howard, W. Klemperer. 
J. Chem. Phys. {\bf 56}, 2442 (1972).
\bibitem{Nemuchin92}
A. V. Nemukhin. 
Zh. Fiz. Khim. {\bf 66}, 4 (1992).
\bibitem{Krachmalnicoff16}
A. Krachmalnicoff,   R. Bounds, S. Mamone, S. Alom, M. Concistre, B. Meier, K. Kouril, M. E. Light, M. R. Johnson,  S. Rols, A. J. Horsewill, A. Shugai, U. Nagel, T. Room, M. Carravetta, M. H. Levitt, R. J. Whitby.
The dipolar endofullerene HF@C60.  Nature Chemistry, {\bf 8}, 953-957 (2016).
\bibitem{Savin20}
A.V. Savin, O.I. Savina.
The Structure and Dynamics of the Chains of Hydrogen Bonds of Hydrogen Fluoride Molecules Inside Carbon Nanotubes.
Physics of the Solid State {\bf 62}(11), 2217-2223 (2020).


\end{references}
\end{document}